# Generating of short pulses with high amplitudes by using of standard Current-Feedback-Amplifier.

Franz Peter Zantis, Dipl.-Ing.(TU)
RWTH Aachen University, Physical Institut IIIa, Electronic Workshop
`zantis@physik.rwth-aachen.de`

The traditional method to generate pulses in the range of some nanoseconds with high amplitudes, is the using of the avalanche-effect of transistors. However the problem with this method is the poor possibility for parameterizing.
That was the trigger to look for a different way: under using of modern electronic components, a pulse generator, which is parameterizable, was developed. The parameters can be set by using a Windows- or Linux-PC. The pulse width can be adjusted in the range of 3....80 ns; the pulse amplitude in the range of 18...84 V and the pulse repetition rate in the range of 1 Hz to 50 kHz.

## 1 Introduction

In physical experiments of particle physics, often short pulses with relatively high energy are necessary. In the current application, a pulser, which is to be developed, must feed an antenna to generate electromagnetic pulses. In other applications it is used to generate ultra-short light pulses with LEDs.

To generate this pulses, the avalanche effect of a transistor can be used. As described e.g. in [1]. But, with this solution, parameterization is not, or only possible with great effort.

A different way is the use of the ECL-technique (details see [9]). With this, parameterization is possible. However, the pulses which comes from an ECL-circuit have very small amplitudes. This means, that an amplifier is mandatory. This is an additional expenditure and an amplifier has maybe not a good influence to the signal shape. But that is not all: the ECL-techniques is also outmoded and the components are not easy to get.

The new upcoming CF-Amplifiers (CFA = Current Feedback Amplifier) with slew rates higher than 6V/ns was the trigger for the idea to generate the pulse inside of the output circuit. This makes the use of an amplification of a generated raw-pulse needless.

## 2 The Idea

The basic idea of this application is to use a Current Feedback Amplifier as comparator. This is unusual, because of the fact, that at minimum one of both inputs of an CFA is a current input. So it must be compared a voltage with a current - which is impossible on the first view.

The concept of the pulse-shaper can be seen in figure 1. A standard microcontroller generates a trigger pulse with individual repetition rate. As described, the CFA works as comparator. The incoming pulse is connected to the positive (voltage-) input (+IN). At this time, the output of the VFA is at low level. So the positive potential at the input +IN of the CFA is higher than the equivalent positive potential at the current input -IN. As result the output of the CFA goes to high-level with its maximum slew-rate and T1 opens. The pulse appears at R7. It is feeded by C2.

The trigger pulse from the microcontroller with the amplitude $\hat{U}_P$ goes also to the delay circuit R1 and C1.

$$u_{C1}(t) = \hat{U}_P \cdot (1 - e^{\frac{-t}{R1 \cdot C1}}) \qquad \{Eq.\ 2.1\}$$

For small $t$ ( $t < RC$), the voltage at C is approximated a linear function (see [2])

$$u_C(t) = \hat{U}_P \cdot (1 - (\frac{-t}{R1 \cdot C1} - 1)) \qquad \{Eq.\ 2.2\}$$

The "slew rate" of the RC-circuit is in this case:

$$sr_{R1C1} = \frac{\hat{U}_P}{R1 \cdot C1} \qquad \{Eq.\ 2.3\}$$

The VFA works also as a comparator. It compares the constant voltage of $U_{ref}$ with the voltage at the condenser C1, which increases with a short delay from the pulse (according to Eq. 2.2). If the amplitude at C1 is higher than $U_{ref}$, the output of the VFA goes to high and changes the voltage level at –IN of the CFA. Here, the slew-rate of the VFA must be considered. Of course, also the slope of the pulse, which comes from the microcontroller, has influence of the delay. So, there are three variables, which have influence to the delay, with which the CFA can be switched.

- the slope of the incoming trigger pulse from the microcontroller $s_{U_P}$

- the "slew rate" of the RC-circuit $sr_{RC}$

- the slew rate of the VFA $sr_{VFA}$

This makes the calculation complicate. Additional, the layout of the circuit board is not free of parasitic capacitive and inductive components. This has e.g. influence on the value of C1.

A very fast VFA needs a bigger capacity of C1. In the best case, if

$$sr_{VFA} \gg sr_{RC}$$

and

$$sr_{VFA} \gg s_{U_P}$$

the slew rate of the VFA can be unconsidered.
On the other hand: with a sufficient slow VFA, the other two variables can be ignored and C1 can be dispensed. However, the question is, if in this case the delay, to toggle the CFA is may be too long to meet the needed pulse width.

If the output of the VFA changes to high, the output of the CFA changes to low level and T1 closes.

Apart from the discussed influence of the slew rates, the pulse width depends on the value of the voltage $U_{ref}$. The higher the value of $U_{ref}$, the wider is the pulse at the output. The values of $U_{ref}$ can be varied in a linear way, between

$$0V < U_{ref} \ll \hat{U}_p$$

Not easy to dimension are the resistors R2, R3 and R5. The resistors form an electrical charge balance for "-IN" as a counterbalance to the potential at "+IN".

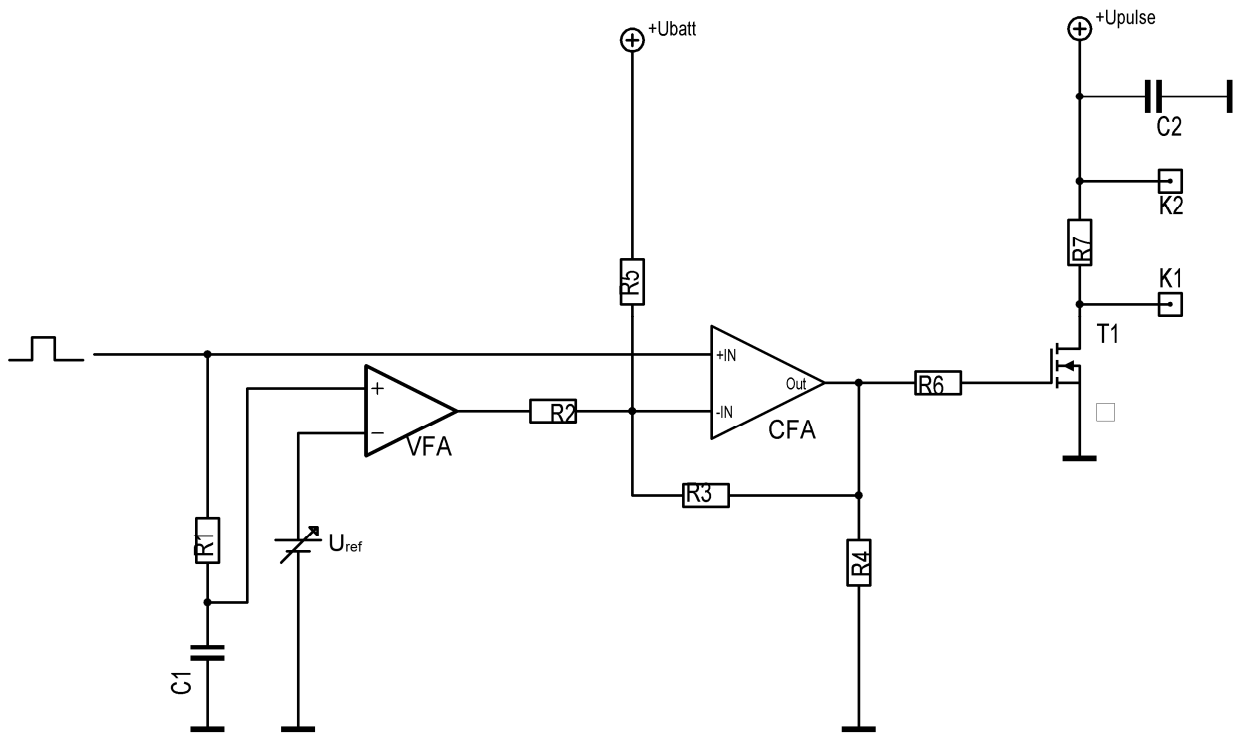

*Fig.1: Concept of the circuit.*

May be in some applications (where amplitudes of about 12V are sufficient) T1 is not necessary and the pulse can be taken immediately from the CFA. Depending on the kind of CFA it can be used in single supply-mode. If T1 is necessary, than it is helpful that the CFA can deliver negative voltage at the output. This will accelerate to switch off T1 by sucking of the charge carriers which are located on the gate of the transistor (see also [8]).

# 3 Selecting the components

With respect on the availability and to limit or lower warehousing, the aim was in this case, to use standard electronic components. The choice fell on the CFA-amplifier THS3202 from Texas Instruments, because of its high slew rate about 9 V/ns. Additional, this component can deliver output currents up to 115 mA and voltages up to 13 V. This is helpful to toggle the FET at the output. Details about the THS3202 can be found in [5].

In fact, the THS3202 is designed to work in HF amplifier. The basic idea here was now, to use this component as a comparator. This mode of operation is not recommended from the manufacturer of this component. They do not support this. However, it works - how this application shows.

For the VFA the type LM6172 is used. The slew rate of this device is 3V/ns. This is fast for a VFA. Details about the LM6172 can be found in [6].

For selecting the FET, the Gate-Source-capacity is a very important point. It must be as small as possible. On the other hand, only a transistor which supports a high drain-source voltage, can supply high output amplitude. Both requirements are contradictory, so that a careful selection of the type is necessary. The type MRFE6VS25NR1 (details see [7]) satisfies both contradictory features for the task very well. With the type, output pulses with amplitudes up to 120V (including a safety distance) are realisable.

If instead of an FET a bipolar transistor should be used, [8] can be helpful to get a short switching time.

# 4 Circuit diagram and design details

The complete circuit diagram of the pulser is shown in fig. 3. In fig. 2 the core circuit of the pulser is shown. The trigger-pulse comes in from a microcontroller. It is amplified with factor 2 by the operational amplifier U4A. The output of U4A goes immediately to the output circuit IC2 (CFA, THS3202). This works as comparator and the output jumps to high. T1 opens and the pulse appears at K1 (Antenna).

It must be considered, that the energy for the pulse cannot come from the power supply, because the pulses are too short. The layout of the PCB, with its parasitic components, makes this impossible. Rather the energy comes from the capacitor C2. It must be placed very close to the output and the transistor T1.

The condenser C2 must be designed for the maximum output voltages (more than 84 V in this case). On the other hand: any parasitic inductance in C2 degrades the output pulse. To avoid inductive components, the capacity should be as small as possible.

Used is a capacity about 100nF. It reaches for impulse durations up to 100 ns if the load has 50 Ω and 2% amplitude-decrease at the end of the pulse is allowed.

$$\frac{U_{C2}}{U} = e^{-\frac{t}{RC}} = e^{-\frac{100ns}{50\Omega \cdot 100nF}} \approx 0,98 \qquad \{\text{Eq. 4.1}\}$$

There are two CFAs in the same case of the THS3202. The second one is used to get an additional amplification of the output-signal if needed. This can be the case, if a different FET is used or the output of the CFA may be used without further parts. The amplification-factor is 2. Depending on which of the CFA should control T1, R26 or R11 is mounted.

The VFA U4B compares the voltage at C4 with a DC-voltage which comes from R27 or from an Digital-to-analogue-converter (DAU). As soon the voltage at C4 is higher as these DC-voltage, the output of U4B changes the current potential at "-IN.A" of the CFA. As result, the output of the CFA IC2 changes from high to low and T1 closes.

The time between opening T1 and closing T1 depends on the time constant of R9/C4 and the slew rate of the VFA U4B. It is possible to work only with the slew-rate of the VFA, if the rest of the circuit can follow.

An important point is, that the CFA IC2 works as comparator – however the negative input "–IN.A" is a current input. It is hard to determine the dimension of the resistors, which are necessary, to get the right working point as comparator. In this case, the resistors R33 and R2 forms together with the internal resistant of IC2 a voltage divider. Whereby the based voltage-potential is determined by the VFA U4B. R25 works as feedback-resistor. With CFAs this resistor is a must to avoid unstable operation. However - it limits the possible smallest pulse width.

All in all, the resistors R33, R2 and R25 form together with the current-input "-IN.A" of the CFA a careful tuned "potential-scale". The values for this resistors have been determined empirically.

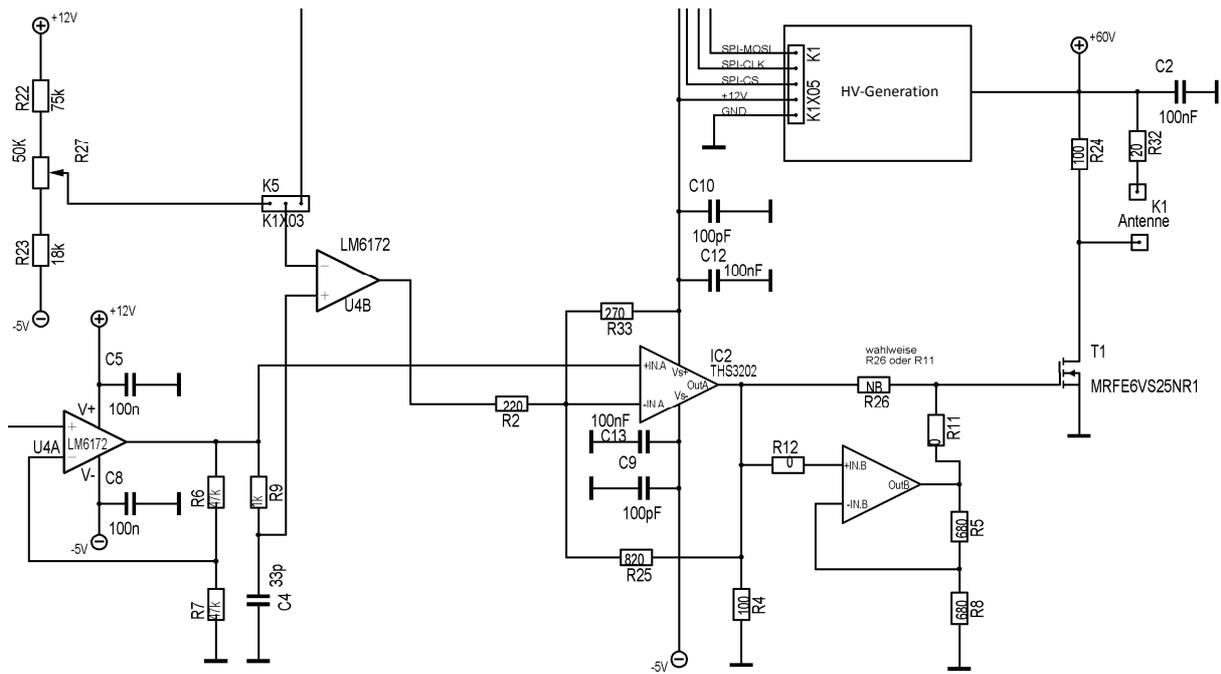

*Fig.2: Core circuit of the pulser.*

Very important is to generate the layout of this circuit very careful. A lot of experience is necessary. The condenser C2 must be placed very close to T1 respectively the output. For designing HF circuits or circuits to handle fast data/pulses, big mass-areas are recommended. However this is not good for CFA. The area around the negative (current-) input mustn't have mass-areas.

The lines between the components should be as short as possible. Especially between the output of the CFA and the gate of T1. Also between the output of U4A and U4B and the input of the CFA. And also between the CFA and the passive components around it (C9, C10, C12, C13, R25, R33, R2).
On the other hand, capacitive feedback between output and input of the THS3202 can result in oscillation. So the components have to be placed close together, but in a way which avoids capacitive feedback.

The output of the CFA changes between negative and positive supply voltage. It is never on 0V. This means, that the CFA must deliver always relative high currents. It is recommended to glue a heatsink on the CFA.

In this application the transistor T1 do not need a heatsink. The transferred power in the moment when the pulse appears at a load of 50 Ω is up to

$$\frac{(84V)^2}{50\Omega} = 141,12W \qquad \{Eq.\ 4.2\}$$

But the duration is not more than 100 ns. The pause until the next pulse, depends on the repetition rate. In this case, the maximum repetition rate is 50 kHz respectively 20 µs. This is 200 times longer than the maximum pulse width. The on-resistance of the transistor is small, but however, if the complete power would be implemented into the transistor, the resulting averaged power is round about

$$\frac{141,12W}{200} \approx 700mW$$

This is the reason, why the transistor does not need a heatsink.

# 5 Controlling the parameters

Controlling the parameters works in this case via a Standard PC and USB-interface. With the help of a Cyrix-processor, which is in the module IO-Warrior56, the USB-protocol is changed into SPI (Serial Peripheral Interface). Details to the SPI-Bus can be found in [10].

Relevant for the pulse width is the voltage at the negative input of U4B. With K5 one can decide if the pulse width should be determined manual via R27 or via a digital-to-analogue converter (which is connected to the SPI-Bus) from the PC.

The repetition rate is controlled by a separate module U1 which gets the desired value from the SPI-bus. This module works with an MPS430F2013-microcontroller which works as generator. It produces pulses with a amplitude of 3,3 V and a constant width of 40 µs. With K6 one can decide if the trigger pulses come from U1 or from an external generator.

The pulse amplitude is determined by the module "HV-Generation". It is an independent working boost converter. The basic concept is similar to an application from [3] and [4]. The value of the output voltage is controlled via the SPI-Bus. In the first application the pulser works to produce electromagnetic pulses via an antenna. For this application the voltage can be changed between 18 and 84 V. the help of an antenna, the value can be changed between 18 and 84 V.

Fig. 4 shows an example pulse of the described pulser. Fig. 5 shows the top side of the prototype and fig. 6 the screen of the control program.

*Fig.3: The complete circuit diagram. This pulser is supplied from an accumulator of a model-helicopter.*

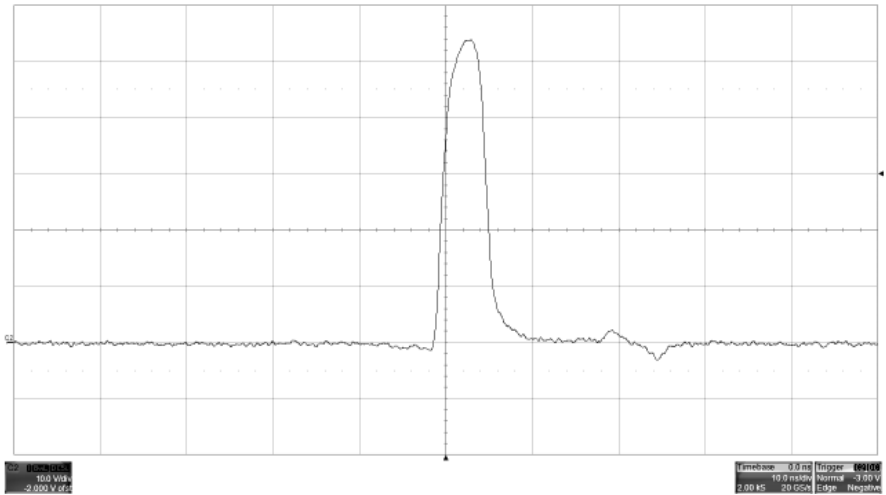

*Fig.4: Example pulse with a width of approximately 5 ns and an amplitude of about 54V. The load was 50Ω.*

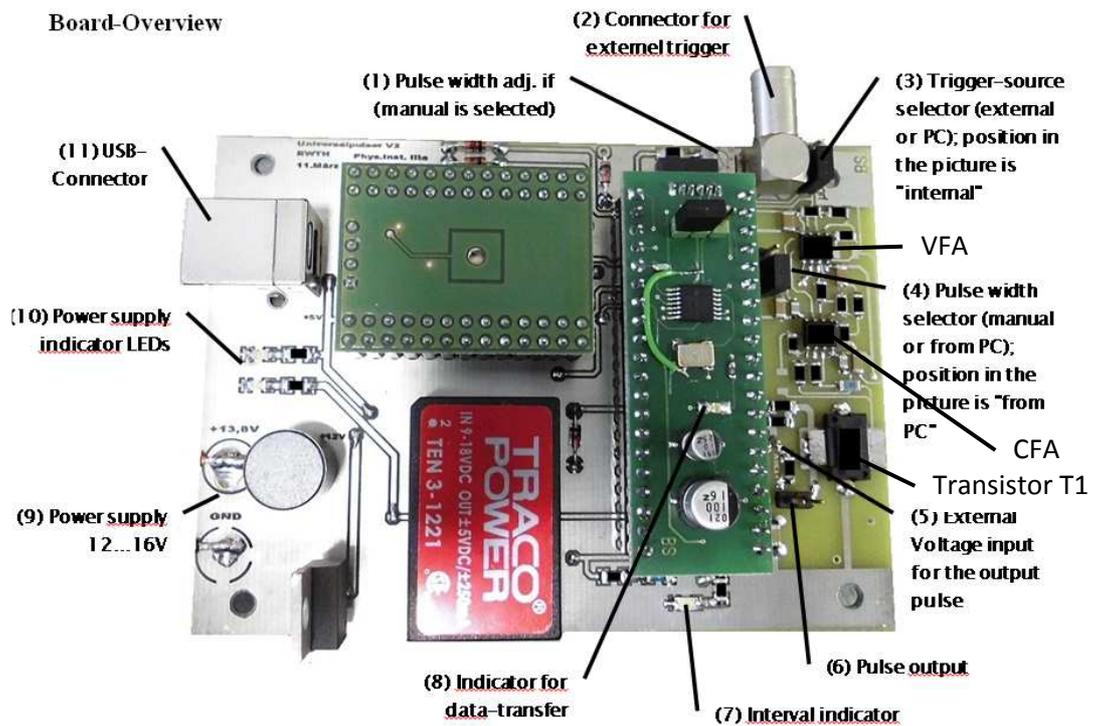

*Fig.5: Prototype, top view (heatsink on the CFA is not mounted).*

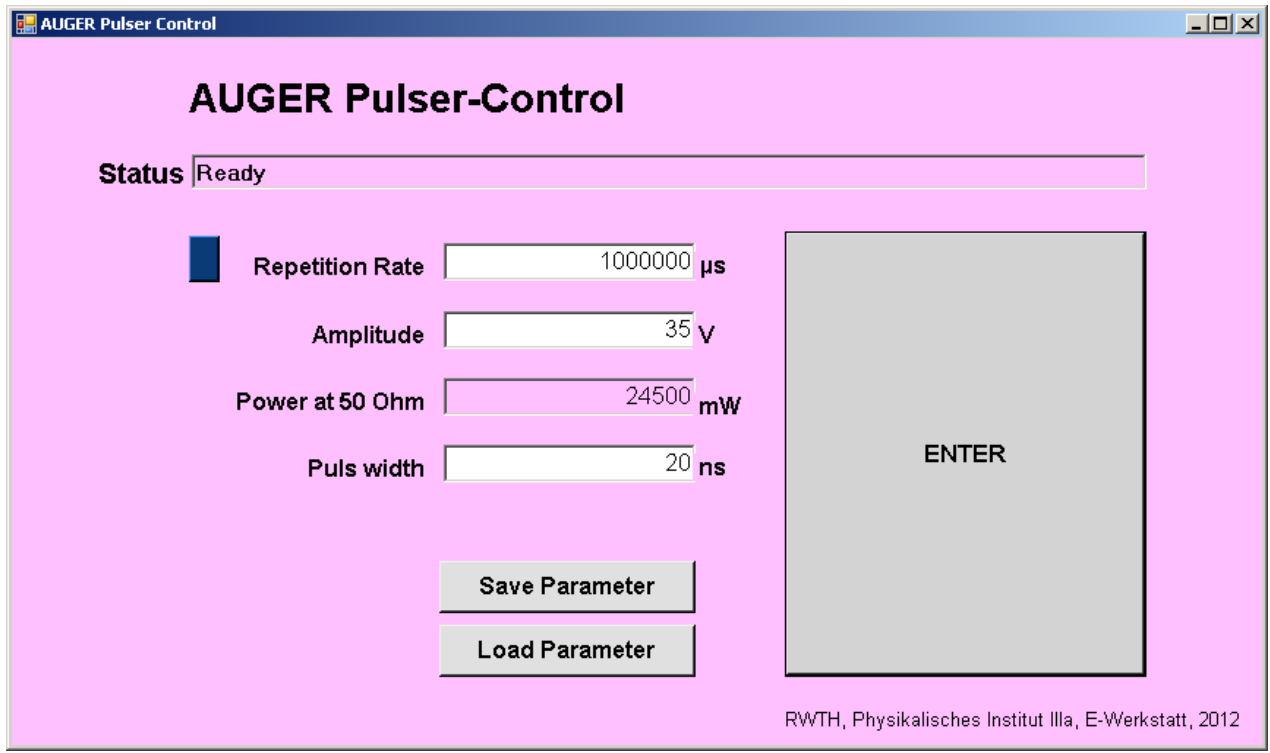

*Fig.6: Screen of the control program.*